\documentclass[12pt]{article}

\usepackage[utf8]{inputenc}

\usepackage[colorlinks,bookmarksopen,bookmarksnumbered,citecolor=red,urlcolor=red]{hyperref}

\usepackage{amsmath} 
\usepackage{amssymb} 
\usepackage{amsfonts}
\usepackage{stmaryrd}

\usepackage{txfonts}

\usepackage{pgf}
\usepgflibrary{arrows.meta}

\newcommand{\xcomm}[1]{{\small\textit{/\kern+0.1em/ #1}}}

\newcommand{\fun}{\ensuremath{\rightarrow}}
\newcommand{\power}{\ensuremath{\mathbb{P}}}
\renewcommand{\emptyset}{\varnothing}
\newcommand{\limg}{\ensuremath{\mathopen{(\kern-0.4ex|}}}
\newcommand{\rimg}{\ensuremath{\mathclose{|\kern-0.4ex)}}}

\newcommand{\xD}{\ensuremath{\mathbf D}}
\newcommand{\xS}{\ensuremath{\mathbf S}}
\newcommand{\xR}{\ensuremath{\mathbf R}}
\newcommand{\xRR}{\ensuremath{\widehat\xR}}
\newcommand{\xSN}{\ensuremath{\xS_{\mathrm N}}}
\newcommand{\xSX}{\ensuremath{\xS_{\mathrm X}}}

\newcommand{\makeobjectTT}[1]%
{\expandafter\newcommand\csname x#1\endcsname{\ensuremath{\mathtt{#1}}}}
\newcommand{\makeobjectIT}[1]%
{\expandafter\newcommand\csname x#1\endcsname{\ensuremath{\mathit{#1}}}}
\newcommand{\makeobjectSF}[1]%
{\expandafter\newcommand\csname x#1\endcsname{\ensuremath{\mathsf{#1}}}}
\newcommand{\makeobjectBF}[1]%
{\expandafter\newcommand\csname x#1\endcsname{\ensuremath{\mathbf{#1}}}}

\makeobjectSF{in}
\makeobjectSF{out}
\makeobjectSF{instances}
\makeobjectSF{bindings}
\makeobjectSF{singletons}
\newcommand{\xxembed}{\ensuremath{\mathsf{embed}}}
\newcommand{\xembed}{\ensuremath{\overline\xxembed}}

\newcommand{\xbox}[4]{
    \pgfnodecircle{#3}[fill]{\pgfxy(#1,#2)}{0.1cm}
    \pgfputat{\pgfxy(#1,#2)}{
      \pgfputat{\pgfxy(0.1,0)}{\pgfbox[center,top]{#4}}}}

\newcommand{\xxbox}[4]{
    \pgfnoderect{#3}[stroke]{\pgfxy(#1,#2)}{\pgfxy(0.4,0.4)}
    \pgfputat{\pgfxy(#1,#2)}{
      \pgfputat{\pgfxy(0.3,0)}{\pgfbox[left,center]{#4}}}}

\newcommand{\xnew}[1]{#1}
\newcommand{\xxnew}[1]{#1}
\newcommand{\xumbi}[1]{#1}

\begin{document}
\title{Critical Semantic Properties\\
  of Music Notation Datasets\\
(Talk held on the Music Encoding Conference 2024, Trenton, TX)}

\author{Markus Lepper \and Baltasar Trancón y Widemann}

\date{}

\maketitle

\begin{abstract}
  The semantics of notation systems can naturally be meta-modelled
  as a network of transformations, starting with the syntactic elements of the
  notation and ending with the parameters of an execution.
  In this context, a digital encoding format for music notation can be
  seen as selecting a subset of the data nodes of this network for
  storage, leaving others to evaluation.
  For such a selection, semantic properties are defined which have
  impact on the practical costs of maintenance, migration, extension, etc.
\end{abstract}

%% keywords: music encoding, data flow network, relational theory
% ---------------------------------------------------------------------------
% \section{Context}

Let in the following the term \emph{model} stand for a mental representation of
some piece of performing arts (composed music, improvised dance, etc.)
Let the term model in a wider sense also stand for a reification of such a model
(printed sheet music, hand-written Laban notation, etc.)

Let \emph{notational model} stand for a model (a) the structure of which
follows explicit syntax rules, and (b) the semantics of which
(i.e.\,the definition of the intended diachronuous piece of art) is
given by a directed, \xnew{contiguous,} and cycle-free network of evaluation rules.\footnote{
These properties are fulfilled by nearly all notation systems found in
practice and held up in the following for simplification, but \xnew{not all} are
mandatory: cycles in the network would require some fixpoint semantics; undirected
edges correspond to constraint programming.
}
This network step by step transforms the elements of the given syntax into
parameters of the intended execution. These syntactic and semantic
rules are called \emph{notation system}. Systematically, a notation system
is a \emph{meta-model} w.r.t.\ its models.
(A video documentation of a dance improvisation is thus a model, but not a
notational model.)

The paradigm  to describe the semantics of a notation system as a transformation pipeline
or network turned out useful both in theory and in practice, see the
toolchains provided by OpenMusic \cite{openmusic} and the module system
used by LMN \cite{lmn}.
It is sensible to define the network in this \emph{forward direction} from
the sheet music to its execution, because (a) this is the original purpose of
all notation, its raison d'être and its historic motivation, and (b) the
calculation results in this direction are \emph{unambiguous} by definition.\footnote{
This does not exclude degrees of freedom in the execution:
``Improvise from this pitch set'' requires that ``this pitch set'' as such is
unambiguously defined.}

% ------------------------------------------------------------------
\begin{figure}[t]
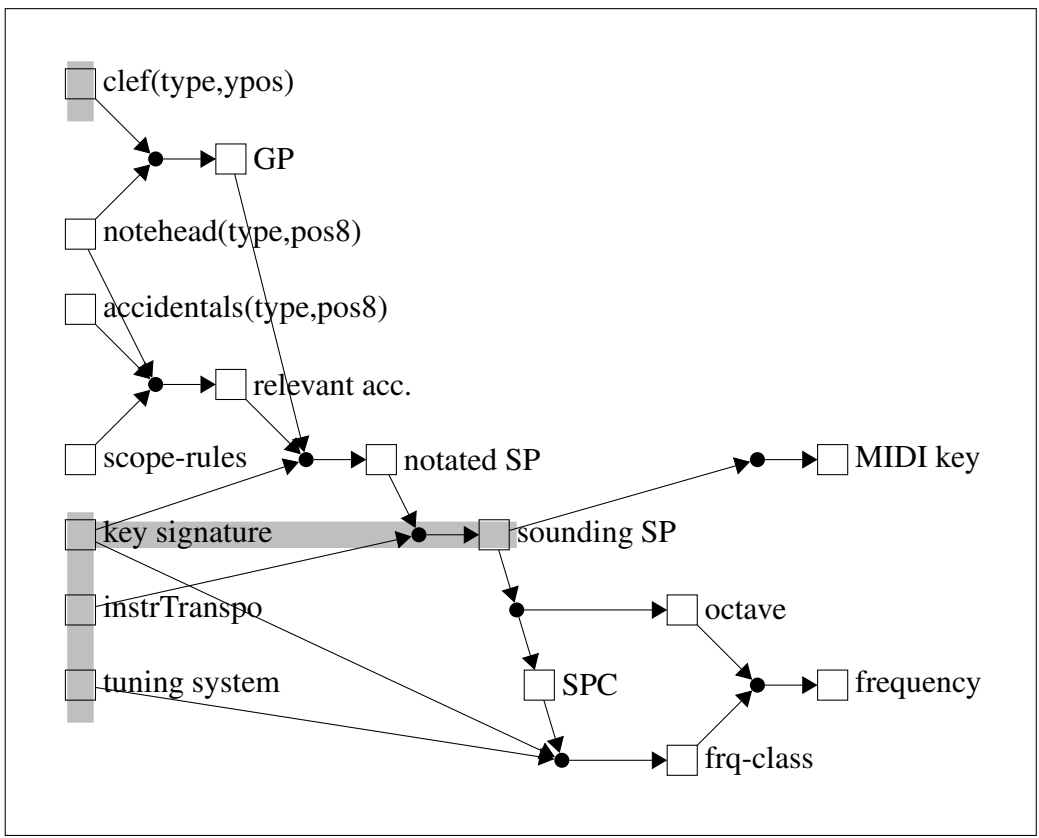

\begin{pgfpictureboxed}{0cm}{4cm}{\textwidth}{15cm}
    \begin{pgfscope}
      \pgfsetlinewidth{10pt}
      \color{lightgray}
      \pgfline{\pgfxy(1.0,13.5)}{\pgfxy(1,14.3)}
      \pgfline{\pgfxy(1.0,5.5)}{\pgfxy(1,8.3)}
      \pgfline{\pgfxy(1,8)}{\pgfxy(6.8,8)}
    \end{pgfscope}

  \pgfsetarrowsend{Triangle[scale=0.3ex]}
   \xxbox{1}{14}{clef}{clef(type,ypos)}
   \xxbox{1}{12}{nh}{notehead(type,pos8)}
   \xbox{2}{13}{gpX}{}
   \xxbox{3}{13}{gp}{GP}
   \pgfnodeconnline{clef}{gpX}
   \pgfnodeconnline{nh}{gpX}
   \pgfnodeconnline{gpX}{gp}

   \xxbox{1}{11}{ssq}{accidentals(type,pos8)}
   \xxbox{1}{9}{scope}{scope-rules}
   \xbox{2}{10}{locaccX}{}
   \xxbox{3}{10}{locacc}{relevant acc.}
   \pgfnodeconnline{nh}{locaccX}
   \pgfnodeconnline{ssq}{locaccX}
   \pgfnodeconnline{scope}{locaccX}
   \pgfnodeconnline{locaccX}{locacc}

   \xbox{4}{9}{npitchX}{}
   \xxbox{5}{9}{npitch}{notated SP}
   \xxbox{1}{8}{key}{key signature}
   \pgfnodeconnline{gp}{npitchX}
   \pgfnodeconnline{locacc}{npitchX}
   \pgfnodeconnline{key}{npitchX}
   \pgfnodeconnline{npitchX}{npitch}
   
   \xbox{5.5}{8}{spitchX}{}
   \xxbox{6.5}{8}{spitch}{sounding SP}
   \xxbox{1}{7}{instr}{instrTranspo}
   \pgfnodeconnline{npitch}{spitchX}
   \pgfnodeconnline{instr}{spitchX}
   \pgfnodeconnline{spitchX}{spitch}

   \xbox{10}{9}{dodec}{}
   \pgfnodeconnline{spitch}{dodec}
   \xxbox{11}{9}{midi}{MIDI key}
   \pgfnodeconnline{dodec}{midi}

   \xbox{6.8}{7}{splitpitch}{}
   \xxbox{9}{7}{octave}{octave}
   \xxbox{7.1}{6}{pclass}{SPC}
   \pgfnodeconnline{spitch}{splitpitch}
   \pgfnodeconnline{splitpitch}{octave}
   \pgfnodeconnline{splitpitch}{pclass}

   \xxbox{1}{6}{tuning}{tuning system}
   \xbox{7.4}{5}{fx}{}
   \xxbox{9}{5}{fc}{frq-class}
   \pgfnodeconnline{pclass}{fx}
   \pgfnodeconnline{tuning}{fx}
   \pgfnodeconnline{key}{fx}
   \pgfnodeconnline{fx}{fc}

   \xbox{10}{6}{add}{}
   \xxbox{11}{6}{f}{frequency}
   \pgfnodeconnline{fc}{add}
   \pgfnodeconnline{octave}{add}
   \pgfnodeconnline{add}f
   
\end{pgfpictureboxed}
\caption{Example interpretation network for pitch information.
  The gray underlay indicates a data set as found in practice.\label{fig-th}}
\end{figure}
% ------------------------------------------------------------------

Figure~\ref{fig-th} shows a typical meta-model from the realm of
conventionally used common western music notation (CWN) concerning pitches.
The square boxes $\square$ stand for data, the
nodes~$\bullet$~for calculations.
Table~\ref{tab-prop} shows the meta-meta-model as mathematical
\emph{relations}, which makes bidirectional evaluation easier.  
Each data box $\square$ appears as a distinct \xnew{set} $S_{i:1\ldots m}$ contained in $\xS$.
\xnew{The union of all values is $T$.
  For a given selection of sets  $X \subset \xS$, any element from $\power T$ which contains
  exactly one value from each of the sets in $X$ and none from the others is
  in  $\xinstances(X)$.}

Each calculation $\bullet$ appears as a distinct
\xnew{relation} $R_{i:1\ldots k}$ in $\xR$.

The boxes in the leftmost column of the Figure, the so-called ``sources'',
where no arrow ends, 
make up the \emph{original notation syntax} and $\xSN$ in the formulas.
The boxes at the rightmost column, the so-called ``sinks'', where no arrow starts, are
the \emph{execution parameters}  $\xSX$.

Figure~\ref{fig-th} is merely for demonstration, details are not relevant
for the following.\footnote{
Remarkably details are that the
distance between note heads and local accidentals is calculated
from their eight-dimensional position as proposed by \cite{lmn},
and that the network
deals with complete pitches, which are explicitly split into
octave register and pitch class only in the last step before applying
the tuning system.
} %footnote
GP, SP, and SPC stand for ``generic pitch'', ``spelled pitch'', and
``spelled pitch class'', as defined by Hentschel et.al.
%% ALT : \cite{unifiedHarm}.
\cite{Hentschel_2022}

% ---------------------------------------------------------------------------
%\section{Classification of Computer Encodings of Music:
%Functional Indepedency.}

With respect to such a semantic network,
the definition of a  \emph{digital encoding format} of music notation
(for storing, transmitting, automated processing, etc.) can be seen as
selecting a particular subset of the data nodes of the network to be stored,
while the others, especially the execution parameters, are left to
evaluation. This selection is called
\emph{(music notation) data set (w.r.t.\ a semantic network)}
in the following. It appears as $\xD$ in Table~\ref{tab-prop}
and an example is indicated by the gray background in Figure~\ref{fig-th}.
A particular combination of values from a data set is called \emph{data set instance}.

The experiences in theory and practice of relational data base construction
have shown that structural properties of the possible data relations
have severe impact on the costs of maintenance, schema extension,
migration, etc.
(See for example the ``normal forms'' of data base schemas, which are
nowadays conditio sine qua non for professional design.)
It is likely that similar properties defined on the semantics of
music notation data sets can be useful.

A special challenge in case of music notation is that the data sets
can be constructed and used for \emph{two different evaluation purposes} with
widely different mathematical properties: towards the execution parameters
(e.g.\ for automated synthesizer realization, called \emph{forward direction}
in the following) or towards the originally intended
notation (e.g.\ for automated sheet music rendering, called \emph{backward}).
\xumbi{\enlargethispage*{3cm}}
Networks and data sets intended for both directions are called \emph{bidirectional}.
\xumbi{\newpage}

% ---------------------------------------------------------------------
\begin{table}
  $\begin{array}{ll}
    \multicolumn2l{\xcomm{universal types and data:}}\\
    \xS = \{S_1,\ldots,S_m\} & \xcomm{all sets of values (\/$\square$ in the Figure)} \\
    \mathsf{disjoint} (S_1,\ldots,S_m) &\\
    \xSN, \xSX \subset \xS  & \xcomm{values in notation (sources)  and execution (sinks)} \\
    T = S_1\uplus\ldots\uplus S_m & \\
    \xD : \power  \xS & \xcomm{selected data set (Gray background in Figure 1)}\\
    \xinstances : \power \xS \fun \power \power T & \xcomm{($\power$ means ``power set''; $\fun$ means ``total function)}\\
     \multicolumn2l{
       \xinstances~ X = \{ v \in \power T \mid
       (\forall s \in \xS \bullet \# (s\cap v) = [s \in X]) \} }\\
     & \xcomm{($[\ldots]$ is Iverson bracket: yields 0/1 for false/true)}\\

    \xR = \{R_1,\ldots,R_k\} & \xcomm{all calculations ($\bullet$ in the Figure)} \\
    \xin, \xout : \xR \fun \power  \xS  & \xcomm{domain and range of calculations} \\
%%    \xdf (R_x) : \xin(R_x) \rel \xout(R_x) & \xcomm{definition of a calculation as relation} \\
%%     R_\_ \subset T \times T & \xcomm{define each calculation as $m$-ary relation} \\

%%     R_{i:1\ldots k} : \power\power T & \xcomm{define each calculation as set of allowed combinations}\\
    \xnew{ R_{i:1\ldots k} \subset \xinstances (\xin R_i \cup \xout R_i) }
        & \xcomm{define each calculation as set of allowed combinations}\\

    %%     r \in R_\_  \land  S_\_ \not\in (\xin R_\_ \cup \xout R_\_) \implies  r \cap S_\_ = S_\_ \\
    %% \multicolumn2l{
    %%   \xfrac{R \in \xR ~~~ r \in R ~~ S \in \xS ~~  S \not\in (\xin R_ \cup \xout R)}
    %%   {r \cap S = S}} \\
    %% \multicolumn2l{
    %%   \forall R \in \xR \bullet \forall r \in R, S \in \xS \setminus \xin R \setminus \xout R
    %%   \bullet r \cap S = S } \\
    %% \xRR = R_1 \cap \ldots \cap R_k\\
    %% \multicolumn2l{
    %%   \xembed~R = \{  r \in R \bullet r \cup \bigcup  \xS \setminus \xin R \setminus \xout R \} }\\
    %% \multicolumn2l{
    %%   \xRR = \xembed(R_1, \xS \setminus \xin R_1 \setminus \xout R_1)  \cap \ldots \cap
    %%          \xembed(R_x, \xS \setminus \xin R_x \setminus \xout R_x)  }\\

  \end{array}$

  $\begin{array}{l}
    \\
    \xRR = \xxembed_{\xin R_1 \cup \xout R_1} R_1   \cap \ldots \cap
      \xxembed_{ \xin R_k \cup \xout R_k} R_k  \\
    %%     \xRR = \xembed~R_1 \cap \ldots \cap \xembed~R_k\\
    \xembed_{\_} \_ ,  \xxembed_{\_} \_  : \power \xS  \times \power \power T \fun \power \power T \\
%%    \xembed_{\{X\} \cup W} V = \xembed_W  \{v \xnew:  V, x : X \bullet v \cup \{x\} \}  \\
    \xembed_{\{X\} \cup W} V = \xembed_W  \{v :  V, x : X \bullet v \cup \{x\} \}  \\
    \xembed_\emptyset V  = V\\
    \xxembed_W V  = \xembed_{\xS \setminus W} V \\
  \end{array}$

  $\begin{array}{l}
    \\
    \xcomm{Properties:}\\
%% \xnew{\mathrm{SCHWAECHER}}~~
%%     \textrm{forward-functional} \Longleftrightarrow
%%     \forall D \in \xinstances~\xD, S \in \xSX \\
%%     \quad\quad\bullet \# (\xxembed_\xD~\{D\} \cap \xRR \cap \xxembed_{\{S\}}~\xinstances \xnew{(\{S\})} ) \leq  1 \\

    \xnew{\textrm{forward-functional} \Longleftrightarrow
    \forall D \in \xinstances~\xD
    }\\
    \xnew{\quad\quad\bullet \# (\xxembed_\xD~\{D\} \cap \xRR \cap \xxembed_{\xSX}~\xinstances\,\xSX  ) \leq  1  }\\

    \textrm{forward-injective} \Longleftrightarrow
    \forall i \in \xinstances~\xSX \\
    \quad\quad\bullet  \# (\xxembed_{\xD} ~\xinstances~\xD \cap \xRR \cap \xxembed_{\xSX}~i  ) \leq 1 \\

    \textrm{forward-surjective} \Longleftrightarrow
    \forall i \in \xinstances~\xSX \\
    \quad\quad\bullet  \# (\xxembed_{\xD} ~\xinstances~\xD \cap \xRR \cap \xxembed_{\xSX}~i  ) \geq 1 \\

\xxnew{
    \textrm{forward-surjective}_{S \in \xSX} \Longleftrightarrow
    \forall i \in \xinstances~(\{S\}) } \\
\xxnew{
    \quad\quad\bullet  \# (\xxembed_{\xD} ~\xinstances~\xD \cap \xRR \cap \xxembed_{\{S\}} ~i  ) \geq 1 } \\

     \textrm{forward-total}
    %%  \Longleftrightarrow
    %% \forall D \in \xinstances~\xD, S \in \xSX \\
    %% \quad\quad\bullet   \# (\xxembed_{\xD}~D \cap \xRR \cap \xxembed_{\{S\}}~\xinstances \xnew{(\{S\})}) \geq 1\\

    \xnew{%% \textrm{forward-total} 
      \Longleftrightarrow
    \forall D \in \xinstances~\xD} \\
    \xnew{\quad\quad\bullet   \# (\xxembed_{\xD}~D \cap \xRR \cap \xxembed_{\xSX}~\xinstances\,\xSX) \geq 1 
    }\\

%%     \textrm{forward-minimal} \Longleftrightarrow
%%     \forall Q\in \xD \bullet\exists \xnew{i \in \xinstances~\xD,}  S \in \xSX \\
%%     \quad\quad\bullet   \xxembed_\xD~i \cap \xRR \cap \xxembed_{\{S\}}~\xinstances \xnew{(\{S\})} \\
%%     \quad\quad\not= \xxembed_{\xD\setminus \{Q\}}~i \cap \xRR \cap \xxembed_{\{S\}}~\xinstances \xnew{(\{S\})}
%% %%                  \not= (i\cup D) \cap \xRR \cap S \\
%%     \\
    \xnew{\textrm{forward-minimal} \Longleftrightarrow
      \forall Q\in \xD \bullet\exists i \in \xinstances~\xD
    } \\
    \xnew{\quad\quad\bullet   \xxembed_\xD~i \cap \xRR \cap \xxembed_{\xSX} ~\xinstances\,\xSX
      }\\
    \xnew{\quad\quad\not= \xxembed_{\xD\setminus \{Q\}}~i \cap \xRR \cap \xxembed_{\xSX}~\xinstances\,\xSX
    }\\
\\
\xcomm{The backward properties are the same, simply
  \xnew{replace} $\xSX$ by $\xSN$.}\\

  \end{array}$
  \caption{Definitions of transformation network, data set, \xnew{and} properties.
    \label{tab-prop}}
\end{table}
% ---------------------------------------------------------------------

%X embed schreiben
%X einbettung r rausziehen
%X (i \D)
% surjective

We found the following properties,
see their formal definitions in Table~\ref{tab-prop}:

\textbf{forward-functional} Every combination of values from the data set
evaluates to at most one particular
%%combination of execution parameters.
\xnew{value in each sink node from $\xSX$.
  As a consequence, each combination evaluates to at most one combination of
  values from all $\xSX$.\footnote{
  \xnew{Normally this is considered an indispensible requirement for any sensible
  semantic network.}
  }
}

\textbf{forward-injective} Every combination of execution parameters
results from only one data set instance.
(This property is normally not met by systems derived from conventional
notation. \xnew{But} the rule known from early dodecaphonic works 
to notate every pitch class with always the same
accidental, even if superfluous in conventional usage,
thus writing always $\natural a$ and $\sharp a$ for a and b-flat, respectively,
aims at this property.)

%% \textbf{forward-surjective} Every combination of execution parameters
%% \xnew{(=~every instance of all $\xSX$)}
%% can result from a data set instance.
%% (This is apparently true for Figure~\ref{fig-th} when restricted to the
%% execution parameter ``Midi key'', but inevitably false for ``frequency''.)

\xxnew{
  The dual construction is \textbf{forward-surjective}
  This means that   every combination of execution parameters 
  can result from a data set instance.
  This general property is hardly sensible in practice:
  For instance in traditional CWN, we should not expect that every mathematically
  possible combination of instrument and clef  (or of clef and pitch) can be represented.

  Instead, the more specific \textbf{forward-surjective in a particular parameter} is
  sensible, which says that all clefs can be represented in \emph{some} combination
  with other parameter values.
}

\textbf{forward-total} Every combination of values from the data set
evaluates to at least one particular
\xnew{value in each sink node from $\xSX$.
  As a consequence, each combination evaluates to at least one instance of all $\xSX$.}
%% combination of execution parameters.
(This implies that contradictory data tuple which ``do not make
sense at all'' cannot appear in a data set instance.)

\textbf{forward-minimal} No value set can be
\xnew{removed from the selected data set $\xD$}
%% taken out of the selected set
without changing the evaluation result of the remaining data.
\xnew{(This property is the most complicated to specify.
  The formula in Table~\ref{tab-prop} says that for each
  set $Q$ removed from the data storage selection $\xD$,
  there is a data sink which changes its values at least for one
  combination of input data.
  This property is frequently violated in bidirectional data sets.)}

The properties in backward direction are defined accordingly. Of course,
the practical consequences of the properties are quite different in both
directions.

\xumbi{\newpage}
The example network and data set in Figure~\ref{fig-th}
are forward-injective.
Without the ``frequency'' output they are forward-surjective;
forward-total only if no pitch outside the midi range can be produced.
They are not forward-minimal due ``clef'', ``key signature', and ``instr Transpo''.
They are backward-total, since every sounding pitch can be notated,
but not injective, because there are notes with variants of accidentals.

A further research question is how the semantic properties (in Table~\ref{tab-prop})
can be inferred from the syntactic properties of the network graph (in Figure~\ref{fig-th}),
how they are related to the operational semantics of evaluating the network,
\xnew{and how the properties can sensibly applied to sub-graphs and to
  operations of graph combination.}

\xxnew{By the way, these properties can possibly also be applied to models instead of
meta-models, that is for discussing one concrete encoding of one particular piece.}
We have applied these properties to selected examples \xxnew{of models and meta-models}
from traditional and digital notation formats; a more systematic analysis of selected
formats is ongoing.

\bibliographystyle{plain}
\bibliography{encoding}

\end{document}